\documentclass[aps,prd,onecolumn,nofootinbib,superscriptaddress]{revtex4}
\usepackage{graphicx}
\pdfoutput=1
\usepackage{epsfig}
\usepackage{amsmath}
\usepackage{amsfonts}
\usepackage{amssymb}
\usepackage{color}
\usepackage{dcolumn}
\usepackage{hyperref}
\usepackage{multirow}
\usepackage{booktabs}
\usepackage{MnSymbol,wasysym}
\usepackage{braket,diagbox}
\usepackage{eurosym}
\usepackage{calrsfs}
\usepackage{subfigure}
\providecommand{\U}[1]{\protect\rule{.1in}{.1in}}

\newcommand{\f}{\begin{equation}}
\newcommand{\ff}{\end{equation}}
\newcommand{\fa}{\begin{eqnarray}}
\newcommand{\ffa}{\end{eqnarray}}

\begin{document}

\baselineskip=0.5cm
\title{Parameter constraints on Horndeski rotating black hole through quasiperiodic oscillations }

\author{Meng-He Wu}
\email{mhwu@njtc.edu.cn} 
\affiliation{School of Physics and Electronic Information Engineering, Neijiang Normal University, Neijiang 641112, China}

\author{Hong Guo}
\email{guohong@ibs.re.kr}
\affiliation{Particle Theory and Cosmology Group, Center for Theoretical Physics of the Universe,
Institute for Basic Science (IBS), Yuseong-gu, Daejeon, 34126, Republic of Korea}

\author{Xiao-Mei Kuang}
\email{xmeikuang@yzu.edu.cn}
\affiliation{ Center for Gravitation and Cosmology, College of Physical Science and Technology, Yangzhou University, Yangzhou 225002, China}

\begin{abstract}
\baselineskip=0.5cm
In this paper, we perform small perturbations around the circular timelike orbit  in the equatorial plane of the Horndeski rotating black hole, and analyze the effects of Horndeski hair on the three fundamental frequencies of the epicyclic oscillations. Since this operation can model  the quasiperiodic oscillations (QPOs) phenomena of the surrounding accretion disc, we then employ the MCMC simulation to fit the theoretical results with three QPO events, including GRO J1655-40, XTE J1859+226 and H1743-322, and constrain the characteristic radius $r$, black hole mass $M$ and spinning parameter $a$, and the Horndeski hair parameter $h$. Our constraint on the Horndeski hair parameter is much tighter than QPOs simulation from the existed accretion models, suggesting slight deviation from classical Kerr black hole. 

\end{abstract}

\maketitle
\tableofcontents

\newpage
\section{Introduction}
One of fascinating topic is the study of phenomena that can distinguish black holes in alternative gravity theories from standard black holes in general relativity (GR), particularly in the strong-gravity regime. The characteristics of such regime are well encoded in the radiation spectrum emitted by accretion flows surrounding these compact objects. A significant portion of this radiation originates from deep within their gravitational fields, offering valuable insights into the near-horizon region. However, since radiation cannot escape from the central region of black holes, observational studies must rely on emissions from their immediate surroundings, such as accretion disks \cite{Bardeen:1972fi}. 

Accretion disks around central compact objects emit soft X-ray continuum radiation, whose frequencies provide a means to infer the innermost disk radius. In principle, this radius coincides with the Innermost Stable Circular Orbit (ISCO), which encodes key information about the nature of the central object. Indeed, as early as the 1970s, \cite{Syunyaev1972} proposed utilizing the fast variability of X-ray flux from accreting matter near the source to investigate geodesic motion in the strong-field region of the accretion disk. This idea gained renewed interest with the discovery of astrophysical phenomena known as Quasi-Periodic Oscillations (QPOs) in the X-ray flux of accreting compact objects. These oscillations, with frequencies reaching up to 450 Hz, are intriguingly close to those expected from bound orbits near the ISCO \cite{Lewinbook, Motta:2016vwf}. QPOs are identified through Fourier analysis of noisy, continuous X-ray data from accretion disks in (micro)quasars, which include systems composed of black holes or neutron stars with companion stars in binary configurations. Based on their frequency range, QPOs are typically classified as high-frequency (HF,$ 0.1-1~kHz$) and low-frequency (LF, $<0.1~kHz$) oscillations \cite{Stella:1997tc,Stella:1999sj}.

HF QPOs are often observed in pairs, consisting of an upper and a lower frequency component, with frequency ratios in black hole microquasars typically clustering around $3:2$ \cite{Kluzniak:2001ar}. Notably, the upper frequency is closely associated with the orbital frequency of test particles moving along the stable circular orbit located at the inner edge of the accretion disk. These oscillations provide crucial insights into the properties of the central compact object and may even offer clues about its origin.
Several theoretical models have been proposed to explain these phenomena. The earliest suggestion that QPOs could serve as probes of strong-field gravity was based on a simple model in which the observed QPO frequencies correspond to the geodesic motion of a test particle \cite{Kluzniak1990}. Given their close connection to the motion of test particles near the ISCO, accurately modeling QPO signals presents a powerful diagnostic tool for studying strong gravitational fields.
In particular, the epicyclic motion of test particles—characterized by their orbital, radial, and latitudinal frequencies—plays a crucial role in modeling and interpreting HF QPOs. These oscillations, in turn, provide constraints on theoretical models and their parameters \cite{Bambi:2012pa,Bambi:2013fea,Maselli:2014fca,Jusufi:2020odz,Ghasemi-Nodehi:2020oiz,
Chen:2021jgj,Allahyari:2021bsq,Deligianni:2021ecz,Deligianni:2021hwt,Jiang:2021ajk,Banerjee:2022chn,
Liu:2023vfh,Riaz:2023yde,Rayimbaev:2023bjs,Abdulkhamidov:2024lvp,Jumaniyozov:2024eah,Xamidov:2025oqx,Jumaniyozov:2025wcs,Guo:2025zca} and references therein.
Moreover, upcoming high-precision observations from Insight-HXMT (Hard X-ray Modulation Telescope) \cite{Lu:2019rru} and the next-generation X-ray time-domain telescope Einstein Probe \cite{yuan2018einstein} are expected to impose stringent constraints on the parameters of central compact objects, further advancing our understanding of strong-gravity regimes.

The Schwarzschild and Kerr spacetimes, as key predictions of GR, serve as natural laboratories for testing gravity in the strong field regime. Recent observations of gravitational waves~\cite{LIGOScientific:2016aoc, LIGOScientific:2018mvr,LIGOScientific:2020aai} and direct imaging of supermassive black holes~\cite{EventHorizonTelescope:2019dse, EventHorizonTelescope:2019ths, EventHorizonTelescope:2019pgp, EventHorizonTelescope:2022wkp, EventHorizonTelescope:2022xqj} have shown remarkable agreement with the predictions of Kerr black holes. Further observations, including those from the Next Generation Very Large Array~\cite{2019clrp.2020...32D} and the Thirty Meter Telescope~\cite{TMT:2015pvw}, are expected to provide deeper insights into the strong gravity regime of black hole spacetimes. These observations offer a crucial opportunity to explore, differentiate, and constrain physically viable black hole solutions that exhibit small deviations from the Kerr metric.
Recently, a hairy rotating black hole solution, incorporating  a logarithmic correction to Kerr metric, was constructed within the framework of Horndeski theory, a modified gravity theory that remains free from Ostrogradski instabilities~\cite{Horndeski:1974wa,Kobayashi:2019hrl}.
Horndeski gravity has garnered significant attention in both cosmological and astrophysical communities due to its potential in explaining the  accelerated expansion of the universe and other intriguing  features (see~\cite{Kobayashi:2019hrl} for a comprehensive review). Additionally,  Horndeski gravity has application in holography \cite{Feng:2015oea,Kuang:2016edj,Cisterna:2017jmv,Filios:2018xvy,Jiang:2017imk,Baggioli:2017ojd,Feng:2018sqm,Wang:2019jyw,Zhang:2022hxl,Bravo-Gaete:2020lzs,Bravo-Gaete:2022lno} and references therein, further enriching its theoretical significance.

Furthermore, the Horndeski framework has been employed to test the no hair theorem of black holes, which states that the isolated black hole in GR is fully characterized by only three parameters: mass, electric charge and angular momentum.  Given the presence of an additional scalar field in the action, a natural question arises, whether black hole solutions with scalar hair can exist within this framework. This inquiry is particularly relevant because, akin to GR, Horndeski theory retains diffeomorphism invariance and yields second-order field equations. A significant development in testing the no-hair theorem within Horndeski gravity was achieved through the study of a specific Horndeski theory with the action~\cite{Babichev:2017guv}
\begin{eqnarray}\label{eq:action0}
S=\int d^4x \sqrt{-g}\big[R+Q_2+Q_3\Box\varphi+Q_4R+Q_{4,\chi}\left((\Box\varphi)^2
-(\nabla^\mu\nabla^\nu\varphi)(\nabla_\mu\nabla_\nu\varphi)\right)\big],
\end{eqnarray}
where $g$ is the determinant of the metric; $R$ is the Ricci scalar; $\chi=-\partial^\mu\varphi\partial_\mu\varphi/2$ is the canonical kinetic term; $Q_i~(i=2,3,4)$ are arbitrary functions of the scalar field $\varphi$ and the kinetic term $\chi$, and $Q_{i,\chi} \equiv \partial Q_{i}/\partial \chi$. This action admits a Horndeski rotating metric  in the Boyer-Lindquist coordinates \cite{Walia:2021emv}
\begin{eqnarray}\label{eq:metric}
ds^2&=&g_{tt} dt^2+g_{rr} dr^2+g_{\theta\theta} d\theta^2+g_{\phi\phi} d\phi^2+2g_{t\phi} dtd\phi\nonumber\\
&=&-\left[1-\frac{2\tilde{M}(r)r}{\Sigma}\right]dt^2+\frac{\Sigma}{\Delta}dr^2+\Sigma d\theta^2+\frac{(r^2+a^2)^2-a^2\Delta\sin^2\theta}{\Sigma}\sin^2\theta d\varphi^2-\frac{4a\tilde{M}(r)r}{\Sigma}\sin^2\theta dtd\phi,
\end{eqnarray}
where
\begin{equation}
\Delta=r^2-2\tilde{M}(r)r+a^2,\quad\Sigma=r^2+a^2\cos^2\theta,\quad \tilde{M}(r)=M-\frac{h}{2}\ln\left(\frac{r}{2M}\right),
\end{equation}
with $a$, $M$ and $h$ the parameters related to the black holes' spin, mass and Horndeski hair, respectively. It is easy to see that this metric is also asymptotically flat and reduces to the Kerr spacetime  as $h\to 0$.

Interesting properties of this rotating hairy black hole, including the Komar conserved quantities, thermodynamics, light deflection, shadow, energy emission and extraction and superradiance, can be found in~\cite{Walia:2021emv,Afrin:2021wlj,Jha:2022tdl,Lei:2023wlt,Donmez:2024lfi}. In particular, the authors of \cite{Afrin:2021wlj} constrained the $(a,~h)$ parameter space relevant to the black hole shadow, such as the angular shadow diameter and the deviation from the Schwarzschild shadow, in light of the constraints from the EHT observations. In particular, in \cite{Zhen:2025nah},  some of us studied the precession orbit, Lense-Thirring (LT) precession as well as periastron precession frequencies of the test particle, and
found that both LT frequency and periastron precession frequency of Horndeski rotating black hole and naked singularity behave differently when the particle's orbit approaches the innermost stable circular orbit. Also, the effects of Horndeski parameter on the two precession frequencies are explored.

Our aim of this paper is to borrow the dynamic of the massive particle to further provide parameters constraints on Horndeski rotating black hole via the observed QPO frequencies. To this end, with the use of Markov Chain Monte Carlo (MCMC) algorithm \cite{Foreman-Mackey:2012any}, we shall fit the
theoretical predictions for the QPO frequencies to the observational X-ray data of GRO J1655-40 \cite{motta2014precise}, XTE J1859+226 \cite{motta2022black}
and H1743-322 \cite{Ingram:2014ara}, respectively, which will provide constraints on the parameters of the Horndeski rotating black hole.

The remaining of this paper is organized as follows. In section \ref{sec:motion}, we analyze the equations of motion for a massive particle around the Horndeski rotating black hole, and
consider the epicyclic oscillation of the particle's circular motion to  investigate the three characterized frequencies in the oscillation. In section \ref{sec:constraint} by employing MCMC method, we use the observational data in three QPOs events to  model parameters. Section \ref{sec:conclusions} contributes to our conclusion and discussion.

\section{ Small perturbation of timelike circular orbit around Horndeski rotating black hole}\label{sec:motion}

We investigate the properties of timelike geodesics in the spacetime of a rotating Horndeski black hole. The dynamics of a test particle are governed by the Lagrangian
\begin{equation}\label{eq:L}
\mathcal{L}=\frac{1}{2} g_{\mu \nu} \dot{x}^\mu \dot{x}^\nu,
\end{equation}
where $\dot{x}^\mu \equiv \frac{d x^\mu}{d \lambda}$, and $\lambda$ represents the affine parameter along the geodesic. Given the stationary and axisymmetry of the metric \eqref{eq:metric}, the spacetime is invariant under translations in the coordinates $(t, \phi)$. Consequently, two conserved quantities arise from these symmetries: the conserved energy $\mathcal{E}$ and the conserved axial angular momentum $L_z$. These conserved quantities are expressed as:
\begin{equation}\label{eq:EL}
\mathcal{E}= -\frac{\partial \mathcal{L}}{\partial \dot{t}} = -g_{t t} \dot{t}-g_{t \phi} \dot{\phi}, \quad L_z = \frac{\partial \mathcal{L}}{\partial \dot{\phi}}   =g_{t \phi} \dot{t}+g_{\phi \phi} \dot{\phi}.
\end{equation}
The expressions for $\dot{t}$ and $\dot{\phi}$ can be derived by solving the system of equations for  $\mathcal{E}$ and $L_z$ in terms of the metric components. The solutions are:
\begin{equation}
\begin{aligned}\label{eq:EL2}
\dot{t} & =\frac{\mathcal{E} g_{\phi \phi}+L_z g_{t \phi}}{g_{t \phi}^2-g_{t t} g_{\phi \phi}}, \\
\dot{\phi} & =-\frac{\mathcal{E} g_{t \phi}+L_z g_{t t}}{g_{t \phi}^2-g_{t t} g_{\phi \phi}}.
\end{aligned}
\end{equation}
To account for the conservation of the rest mass of the test particle, the normalization condition for the timelike geodesic must be satisfied:
\begin{equation}\label{eq:U}
 g_{\mu \nu} \dot{x}^\mu \dot{x}^\nu=-1.
\end{equation}
By substituting the metric  \eqref{eq:metric} and  the expressions for $\dot{t}$ and $\dot{\phi}$ from Eq.~\eqref{eq:EL2} into \eqref{eq:U},  the radial and polar components of the motion can be isolated, yielding the following equation:
\begin{equation}\label{eq:U2}
g_{r r} \dot{r}^2+g_{\theta \theta} \dot{\theta}^2=V_{\mathrm{eff}},
\end{equation}
where $V_{\mathrm{eff}}$   represents the effective potential and is given by:
\begin{equation}\label{eq:veff2}
V_{\mathrm{eff}}=\frac{\mathcal{E}^2 g_{\phi \phi}+2 \mathcal{E} L_z g_{t \phi}+L_z^2 g_{t t}}{g_{t \phi}^2-g_{t t} g_{\phi \phi}}-1.
\end{equation}
For further analysis, we consider the circular geodesics of a particle at $r=r_0$ on the equatorial plane($\theta = \pi/2$). In this scenario, the effective potential must satisfy the following conditions:
\begin{equation}\label{eq:Veff}
V_{\mathrm{eff}}(r_0, \pi / 2)=0,  \quad \partial_r V_{\mathrm{eff}}(r_0, \pi / 2)=0, \quad \partial_\theta V_{\mathrm{eff}}(r_0, \pi / 2)=0.
\end{equation}
The angular velocity of the particle as measured by an observer at infinity is given by:
\begin{equation}\label{eq:omga_phi}
\Omega_\phi=\frac{d \phi}{d t}=-\left.\frac{\mathcal{E} g_{t \phi} +  L_z g_{t t}}{\mathcal{E}g_{\phi \phi} +  L_z g_{t \phi}  }\right|_{r=r_0, \theta=\pi / 2}.
\end{equation}
By imposing the conditions specified in Eq.~\eqref{eq:Veff},  the conserved energy $\mathcal{E}$ and angular momentum $L_z$ of the particle can be expressed as:
\begin{equation}
\begin{aligned}\label{eq:ELz}
E & =-\left.\frac{g_{t t}+g_{t \phi} \Omega_\phi}{\sqrt{-g_{t t}-2 g_{t \phi} \Omega_\phi-g_{\phi \phi} \Omega_\phi^2}}\right|_{r=r_0, \theta=\pi / 2}, \\
L_z & =\left.\frac{g_{t \phi}+g_{\phi \phi} \Omega_\phi}{\sqrt{-g_{t t}-2 g_{t \phi} \Omega_\phi-g_{\phi \phi} \Omega_\phi^2}}\right|_{r=r_0. \theta=\pi / 2}.
\end{aligned}
\end{equation}
Furthermore, the angular velocity in Eq. \eqref{eq:omga_phi} can be rewritten as:
\begin{equation}\label{eq:omga_phi2}
\Omega_\phi= \left. \frac{-g_{t \phi, r} \pm \sqrt{\left(g_{t \phi, r}\right)^2-g_{t t, r} g_{\phi \phi, r}}}{g_{\phi \phi, r}} \right|_{r=r_0, \theta=\pi / 2}.
\end{equation}
The sign $\pm$ corresponds to prograde and retrograde orbits, respectively. Prograde orbits refer to trajectories where the angular momentum is aligned with the spin of the black hole, while retrograde orbits have angular momentum antiparallel to the spin. In this work, we focus exclusively on prograde orbits.

In order to model the QPO phenomena of accretion disk by the three fundamental frequencies of the massive particle orbiting the cental object, we introduce small perturbations around the circular orbit in equatorial plane. The perturbed coordinates are expressed as:
\begin{equation}\label{eq:per}
r(t)=r_0+\delta r(t), \quad \theta(t)=\frac{\pi}{2}+\delta \theta(t),
\end{equation}
where $\delta r(t)$ and $\delta \theta(t)$ are small perturbations from the circular orbit. These perturbations are governed by the following equations of motion:
\begin{equation}\label{eq:perEq}
\frac{d^2 \delta r(t)}{d t^2}+\Omega_r^2 \delta r(t)=0, \quad \frac{d^2 \delta \theta(t)}{d t^2}+\Omega_\theta^2 \delta \theta(t)=0,
\end{equation}
where $\Omega_r$ and $\Omega_\theta$ can be expressed as \cite{Ryan:1995wh,Doneva:2014uma}:
\begin{equation}
\begin{aligned}\label{eq:Omega}
& \Omega_r= \left.  \left\{\frac{1}{2 g_{r r}}\left[X^2 \partial_r^2\left(\frac{g_{\phi \phi}}{g_{t t} g_{\phi \phi}-g_{t \phi}^2}\right)-2 X Y \partial_r^2\left(\frac{g_{t \phi}}{g_{t t} g_{\phi \phi}-g_{t \phi}^2}\right)+Y^2 \partial_r^2\left(\frac{g_{t t}}{g_{t t} g_{\phi \phi}-g_{t \phi}^2}\right)\right]\right\}^{1 / 2}   \right|_{r=r_0, \theta=\pi / 2}, \\
& \Omega_\theta=  \left. \left\{\frac{1}{2 g_{\theta \theta}}\left[X^2 \partial_\theta^2\left(\frac{g_{\phi \phi}}{g_{t t} g_{\phi \phi}-g_{t \phi}^2}\right)-2 X Y \partial_\theta^2\left(\frac{g_{t \phi}}{g_{t t} g_{\phi \phi}-g_{t \phi}^2}\right)+Y^2 \partial_\theta^2\left(\frac{g_{t t}}{g_{t t} g_{\phi \phi}-g_{t \phi}^2}\right)\right]\right\}^{1 / 2}  \right|_{r=r_0, \theta=\pi / 2},
\end{aligned}
\end{equation}
where the quantities $X$ and $Y$ are defined as:
\begin{equation}\label{eq:XY}
X=g_{t t}+g_{t \phi} \Omega_\phi, \quad Y=g_{t \phi}+g_{\phi \phi} \Omega_\phi.
\end{equation}
Finally, the orbital frequency $\nu_\phi$, the radial epicyclic frequency $\nu_r$ and vertical epicyclic   frequency $\nu_\theta$, expressed in standard physical units, are given by:
\begin{equation}\label{eq:nu}
\nu_i=\frac{1}{2 \pi} \frac{c^3}{G M} \Omega_i  \quad  \quad (i= r,\theta, \phi).
\end{equation}

\begin{figure}[ht]
    \centering
    \includegraphics[width=0.39\linewidth]{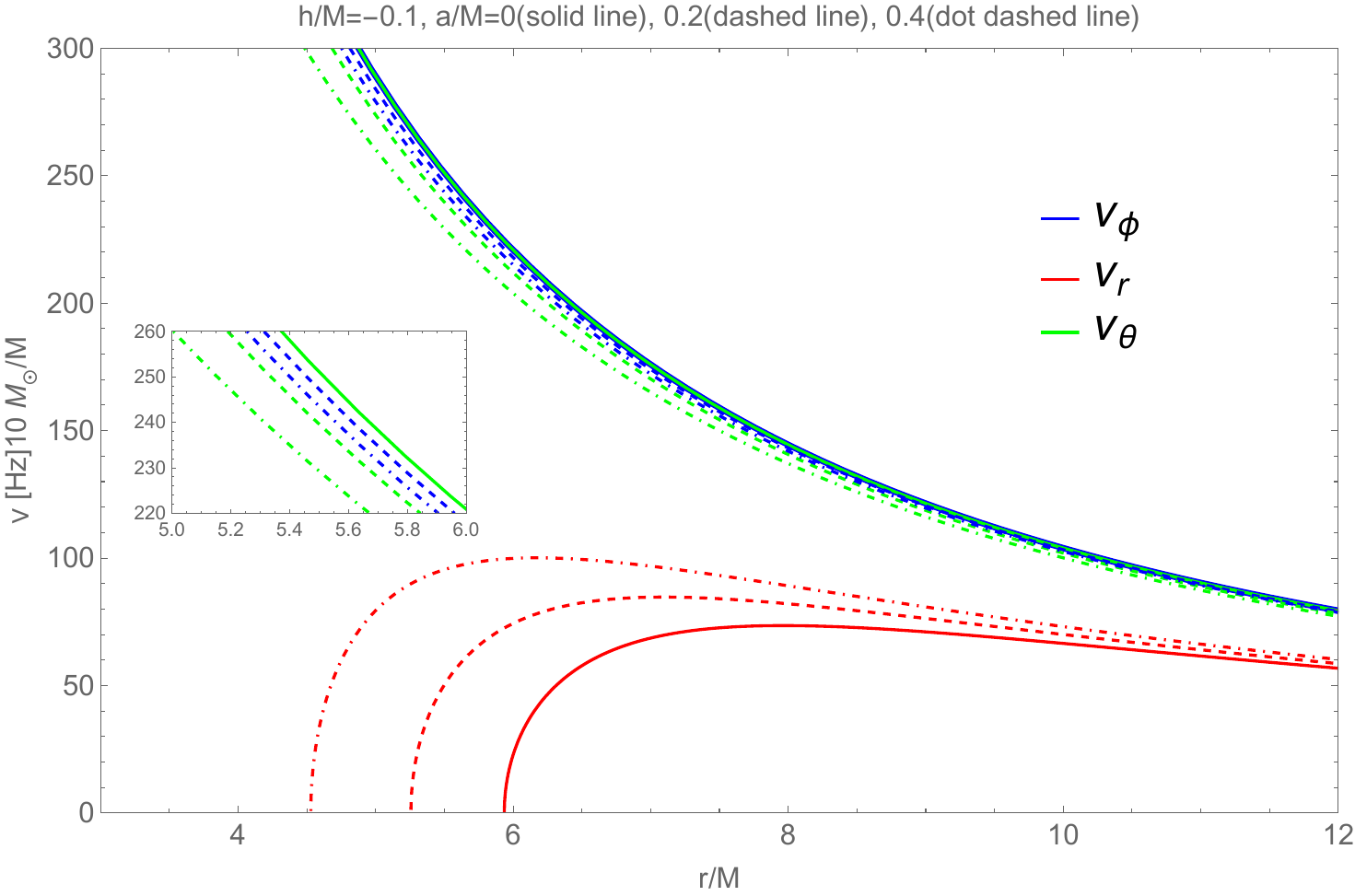}\hspace{1cm}
       \includegraphics[width=0.39\linewidth]{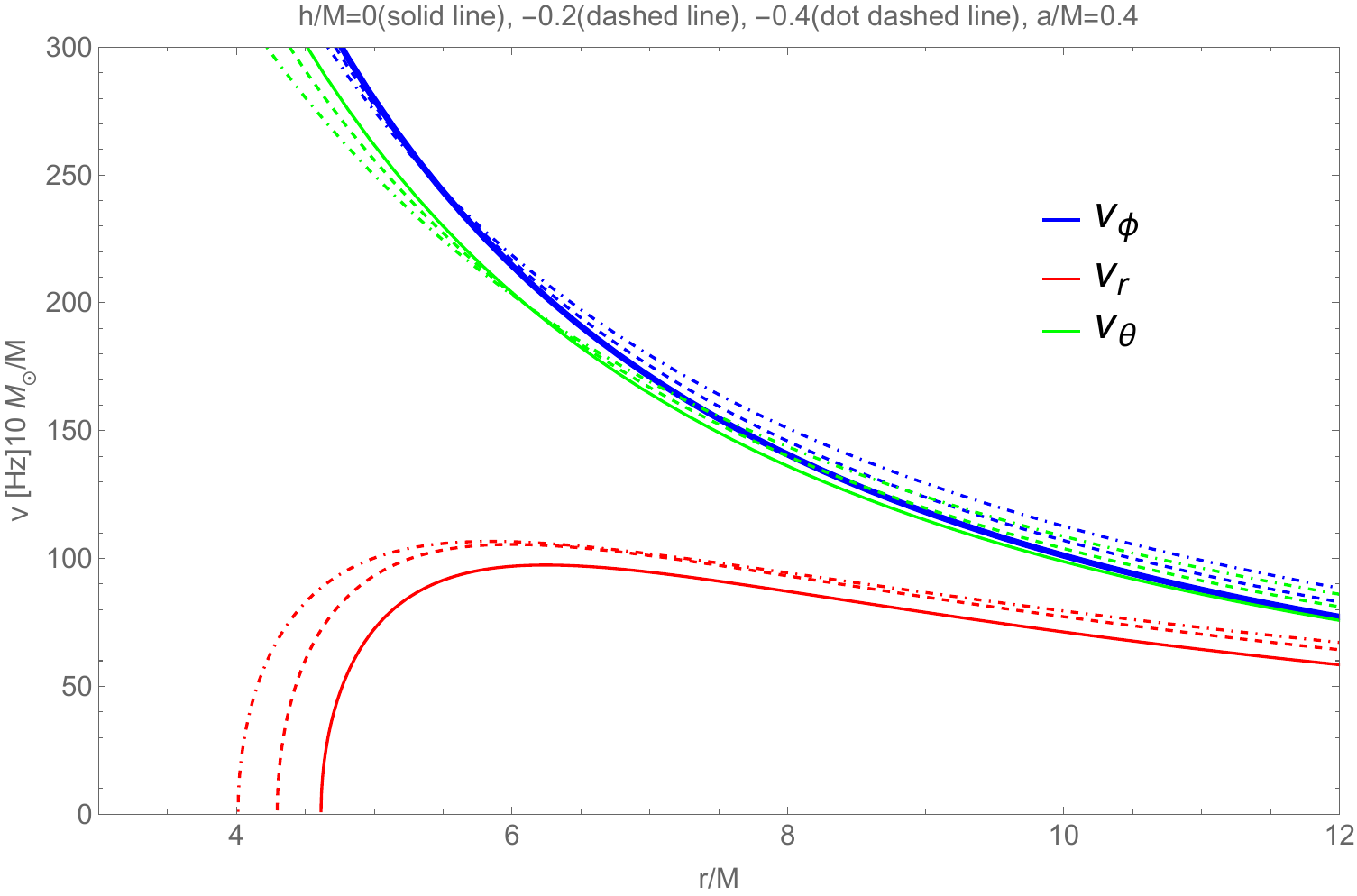}\hspace{1cm}
    \caption{The orbital frequency $\nu_\phi$, the radial epicyclic frequency $\nu_r$ and the vertical epicyclic frequency $\nu_\theta$ as functions of ratio $r/M$ for various values of $a/M$ and $h/M$.}
    \label{fig3}
\end{figure}

As illustrated in Fig.~\ref{fig3}, we present the relationship between the orbital frequency $\nu_\phi$, the radial epicyclic frequency $\nu_r$, the vertical epicyclic frequency $\nu_\theta$, and the ratio $r/M$ for various values of $a/M$ and $h/M$. The plot on the left indicates that when $a/M = 0$, $\nu_\phi$ and $\nu_\theta$ coincide and both frequencies decrease as $a/M$ increases, with the decline of $\nu_\theta$ being more pronounced than that of $\nu_\phi$. Additionally, $\nu_r$ shifts toward smaller values of $r$ as $a/M$ increases.
As illustrated on the right side of the plot, in the relatively small range of $ r $, both $ \nu_\phi $ and $ \nu_\theta $ decrease as the parameter $ h/M $ increases.  Conversely, in the relatively large range of $ r $, both $ \nu_\phi $ and $ \nu_\theta $ exhibit an increasing trend with increasing $ h/M $. In particular, for $ \nu_\phi $, the curves corresponding to different values of $ h/M $ intersect at a specific point, as indicated by the orange dot in the illustration. Furthermore, $ \nu_r $ also shifts towards smaller values $ r $ as $ h/M $ increases.

\section{ Parameter constraints by X-ray observations  of  quasiperiodic oscillations}\label{sec:constraint}

In this section, we present an analysis of QPOs observed in X-ray binaries, including GRO J1655-40, XTE J1859+226 and H1743-322, to constrain the parameters of rotating black holes described by the Horndeski theory. The processed QPO events are utilized, with their frequencies listed in Table~\ref{Table1}. We introduce the periastron precession frequency, $\nu_{\text{per}}$ and the nodal precession frequency, $\nu_{\text{nod}}$, following \cite{stella1999correlations}, defined as: 
\begin{equation}\label{eq:pernod}
\nu_{p e r}=\nu_\phi-\nu_r, \quad \nu_{n o d}=\nu_\phi-\nu_\theta .
\end{equation}
It is noted that $\nu_{p e r}$ and $\nu_{n o d}$ describe the precession of the orbital plane and the precession of the test particle’s orbit, respectively.  Their properties in natural unit for Horndeski rotating black hole and naked singularity have been discussed in \cite{Zhen:2025nah}, which found to be different from those in Kerr black hole.
Here, we shall combine the theoretical results and observational data of QPOs, and employ MCMC simulation to explore the space of physical parameters and to constrain the range of the parameters $a/M$ and $h/M$ of the Horndeski rotating black hole.

\begin{table}[h]
  \centering
\renewcommand{\arraystretch}{1.5}
  \caption{The mass, orbital frequency $\nu_\phi$, the periastron precession frequency $\nu_{\text{per}}$ and the nodal precession frequency $\nu_{\text{nod}}$ of QPOs from the X-ray Binaries selected for analysis.}
  \label{Table1}
\begin{tabular}{ccccc}
\hline \hline
Parameter & $M\left(M_{\odot}\right)$ & $\nu_\phi(\mathrm{Hz})$ & $\nu_{\text{per}}(\mathrm{Hz})$ & $\nu_{\text{nod}}(\mathrm{Hz})$ \\
\hline
GRO J1655-40 & $5.4 \pm 0.3$ \cite{motta2014precise} & $441 \pm 2$ \cite{motta2014precise} & $298 \pm 4$ \cite{motta2014precise} & $17.3 \pm 0.1$ \cite{motta2014precise} \\
XTE J1859+226 & $7.85 \pm 0.46$ \cite{motta2022black} & $227.5_{-2.4}^{+2.1}$ \cite{motta2022black} & $128.6_{-1.8}^{+1.6}$ \cite{motta2022black} & $3.65 \pm 0.01$ \cite{motta2022black}\\
H1743-322 & $\gtrsim 9.29$ \cite{Ingram:2014ara} & $240 \pm 3$ \cite{Ingram:2014ara}  & $165_{-5}^{+9}$ \cite{Ingram:2014ara} & $9.44 \pm 0.02$ \cite{Ingram:2014ara} \\
\hline \hline
\end{tabular}
\end{table}

\subsection{The Monte Carlo Markov chain simulation}

This subsection derives constraints on rotating black holes using the $emcee$ algorithm from \cite{foreman2013emcee}. By Bayes' theorem, the posterior probability of model parameters $(\Theta)$, given the observed data $(\mathcal{D})$, is:
\begin{equation}\label{eq:Pos}
\mathcal{P}(\Theta \mid \mathcal{D})=\frac{P(\mathcal{D} \mid \Theta) P(\Theta )}{P(\mathcal{D} )} ,
\end{equation}
where $P(\mathcal{D} \mid \Theta)$ is the likelihood of the data given the model, $P(\Theta)$ denotes the prior distribution over the parameters, and $P(\mathcal{D})$ is the evidence, serving as the normalizing factor.
In our scenario, $\mathcal{D}$ represents the QPO frequencies for each X-ray binary, while $\Theta$ represents the astrophysical parameters involved in the QPOs events.
The priors are Gaussian within boundaries, i.e. $P(\mu_i) \sim \exp\left[-\frac{1}{2}\left(\frac{\mu_i - \mu}{\sigma_i}\right)^2\right]$, where $\mu_{\text{low},i} < \mu_i < \mu_{\text{high},i}$ for parameters $\mu_i = [M, a/M, r/M]$ with corresponding $\sigma_i$.  For $h/M$, we use a uniform prior, $P(h) = 1$ for $h \in [h_{\text{low}}, h_{\text{high}}]$, and $P(h) = 0$ otherwise. 
The prior values for the Horndeski rotating black hole parameters are taken from Table~\ref{Table2} \cite{Liu:2023vfh,Liu:2023ggz}.

\begin{table}[h]
\centering
\renewcommand{\arraystretch}{1.5}
\caption{The Gaussian prior of the Horndeski rotating BH from QPOs for the X-ray Binaries.}
\label{Table2}
\begin{tabular}{cc@{\hspace{1.0cm}}c@{\hspace{1.0cm}}c@{\hspace{1.0cm}}c@{\hspace{1.0cm}} c}
\hline \hline 
Parameter &  & $M\left(M_{\odot}\right)$ & $a/M$ & $r / M$ & $h/M$\\ 
\hline
\multirow{2}{*}{GRO J1655-40} 
& $\mu$    & 5.307  & 0.286 & 5.677 & \multirow{2}{*}{Uniform $( -1, 0]$} \\ 
& $\sigma$ & 0.066  & 0.003 & 0.035 & \\ 
\multirow{2}{*}{XTE J1859+226} 
& $\mu$    & 7.85   & 0.149 & 6.85  & \multirow{2}{*}{Uniform $( -1, 0]$} \\ 
& $\sigma$ & 0.46   & 0.005 & 0.18  & \\ 
\multirow{2}{*}{H1743-322} 
& $\mu$    & 9.29   & 0.27  & 5.55  & \multirow{2}{*}{Uniform $( -1, 0]$} \\ 
& $\sigma$ & 0.46   & 0.013 & 0.22  & \\ 
\hline \hline 
\end{tabular}
\end{table}

Using the orbital, periastron precession and nodal precession frequencies from Eqs.~\eqref{eq:nu} and \eqref{eq:pernod}, three data sets are used in the MCMC analysis. The likelihood function $\mathcal{L}$ is given by:
\begin{equation}\label{eq:L}
\log \mathcal{L}=-\frac{1}{2} \sum_i \frac{\left(\nu_{\phi, \mathrm{obs}}^i-\nu_{\phi, \mathrm{th}}^i\right)^2}{\left(\sigma_{\phi, \mathrm{obs}}^i\right)^2}-\frac{1}{2} \sum_i \frac{\left(\nu_{per, \mathrm{obs}}^i-\nu_{per, \mathrm{th}}^i\right)^2}{\left(\sigma_{per, \mathrm{obs}}^i\right)^2}-\frac{1}{2} \sum_i \frac{\left(\nu_{nod, \mathrm{obs}}^i-\nu_{nod, \mathrm{th}}^i\right)^2}{\left(\sigma_{nod, \mathrm{obs}}^i\right)^2},
\end{equation}
where $\nu^{i}_{\phi,\text{obs}}$, $\nu^{i}_{per,\text{obs}}$ and $\nu^{i}_{nod,\text{obs}}$ are the observed orbital, periastron precession and nodal precession frequencies, respectively, with theoretical predictions $\nu^{i}_{\phi,\text{th}}$, $\nu^{i}_{per,\text{th}}$ and $\nu^{i}_{nod,\text{th}}$. The uncertainties $\sigma^{i}_{x,\text{obs}}$ represent the statistical errors for each quantity.

We perform MCMC to constrain the parameters $ \{ M, a/M, r/M, h/M \}$ of the Horndeski rotating black hole. Gaussian priors are based on parameter values from the literature. Using the Gaussian prior distribution, $10^5$ random samples are generated for each parameter, covering the physically allowed space within the set boundaries, to determine the best-fit values.

\subsection{Results}

\begin{figure}[h]
    \centering
    \subfigure[~GRO J1655-4]
    {\includegraphics[width=0.7\linewidth]{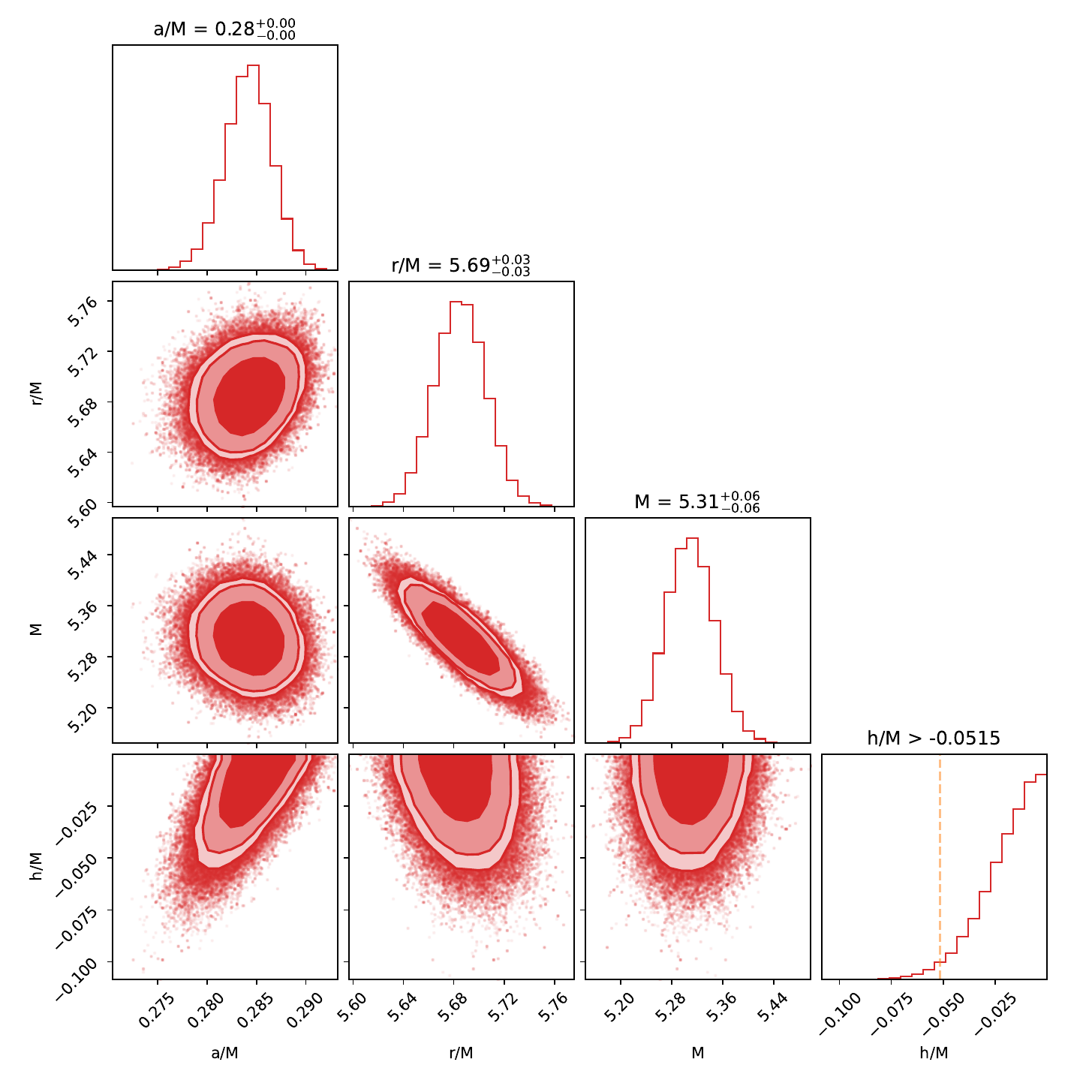}}
     \subfigure[~XTE J1859+226]  
    {\includegraphics[width=0.48\linewidth]{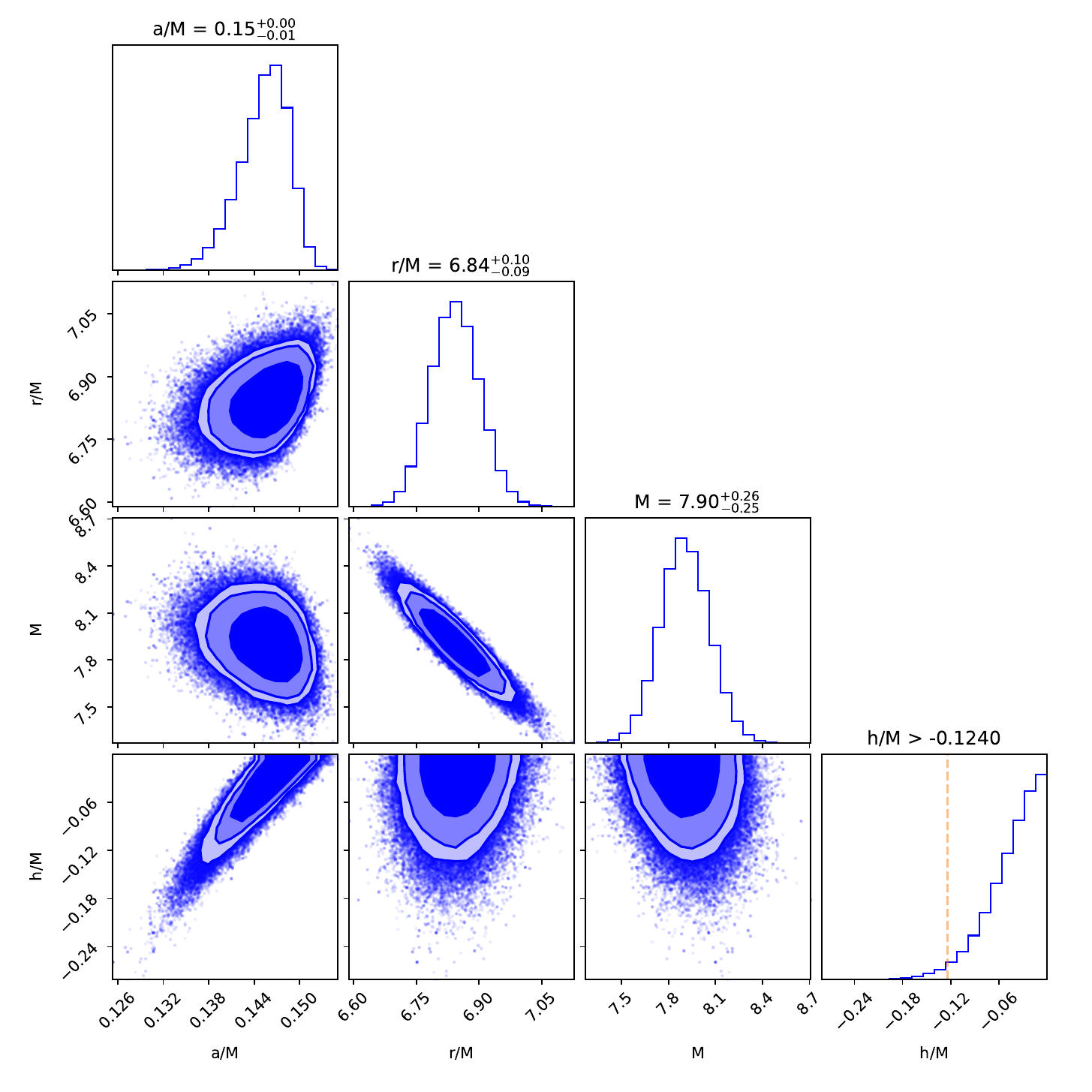}}
   \subfigure[~H1743-322]   
     {\includegraphics[width=0.48\linewidth]{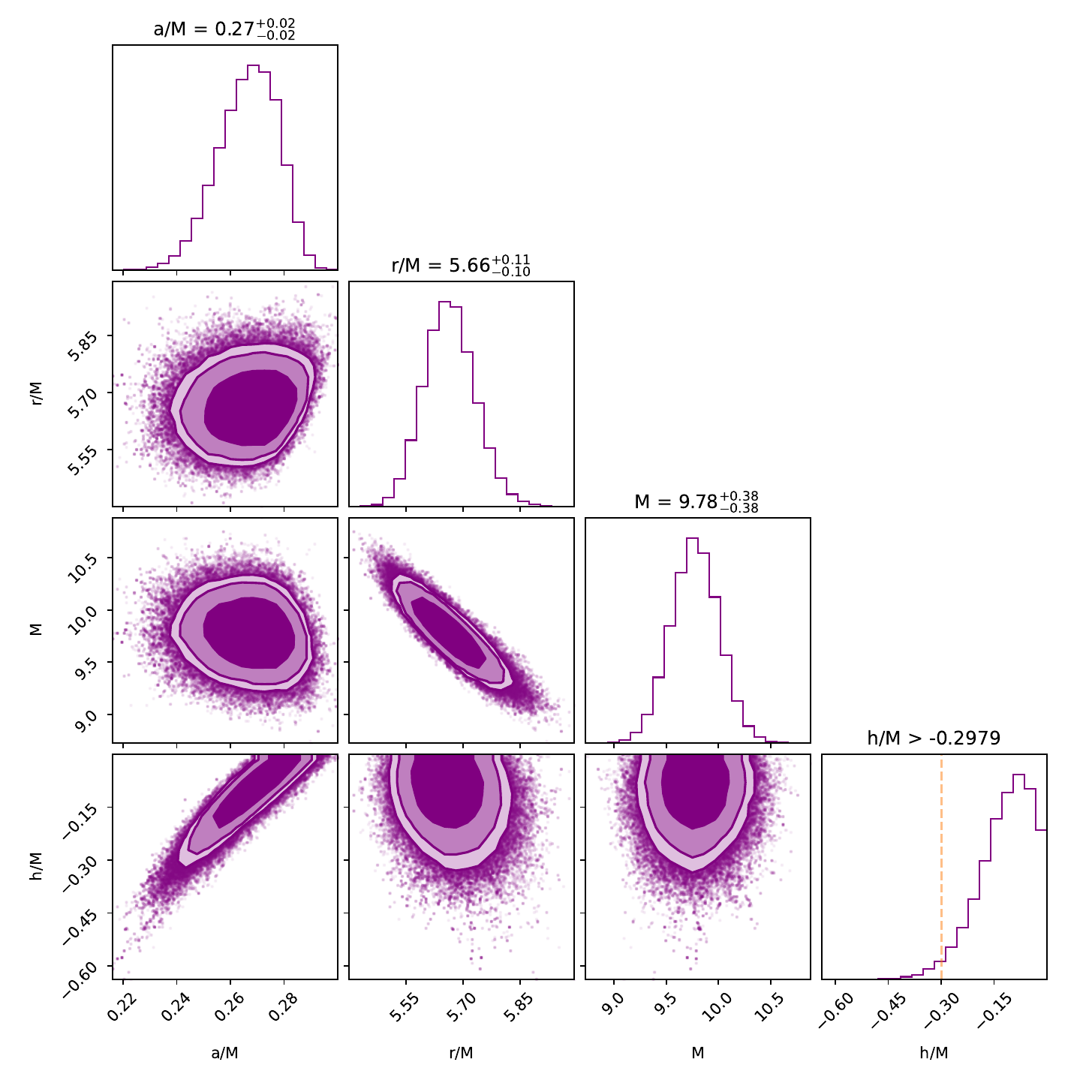}}
    \caption{Constraints on the parameters of the Horndeski rotating black hole with GRO J1655-4, XTE J1859+226 and  H1743-322 from current observations of QPOs within the relativistic precession model.}
    \label{fig-corner}
\end{figure}

In this subsection, we present the four parameters of the Horndeski rotating black hole, derived from MCMC analysis. The best-fitting values for these parameters are provided in Table~\ref{Table3}. In Fig~\ref{fig-corner}, we illustrate the MCMC analysis results for all parameters of the Horndeski rotating black hole during three X-ray QPO events. In the contour plots of these figures, the shaded regions reveal the $68 \%$, $ 90 \%$ and $95 \%$ confidence levels (C.L.) of the posterior probability density distribution for the complete parameter set.

\begin{table}[h]
  \centering
  \renewcommand{\arraystretch}{1.5}
  \caption{The best-fit values of the Horndeski rotating black hole parameters from QPOs for the X-ray Binaries.}
  \label{Table3}
\begin{tabular}{c@{\hspace{1.0cm}}c@{\hspace{1.0cm}}c@{\hspace{1.0cm}}c@{\hspace{1.0cm}}c}
\hline \hline
Parameter     & $M\left(M_{\odot}\right)$ & $a/M$ & $r/M$ & $h/M$\\
\hline
GRO J1655-40  & $5.31 ^{+0.06}_{-0.06}$    & $0.28 ^{+0.00}_{-0.00}$ & $5.68 ^{+0.03}_{-0.03}$    & $>-0.0509$ \\
XTE J1859+226 & $7.90^{+0.26}_{-0.25}$          & $0.15^{+0.00}_{-0.01}$ & $6.85 ^{+0.10}_{-0.10}$   & $>-0.1250$ \\
H1743-322     & $9.78 ^{+0.38}_{-0.37}$ & $0.27^{+0.02}_{-0.02} $& $5.66^{+0.11}_{-0.10}$   & $>-0.2943$ \\
\hline \hline
\end{tabular}
\end{table}

Table~\ref{Table3} presents the spin parameters $ a/M $ for the three systems, ranging from $0.15$ to $0.28$. This variation suggests that differing rotation speeds of black holes significantly affect QPQs. The spin values of $ a/M \approx 0.28 $ for GRO J1655-40 and H1743-322 could result in enhanced frame dragging, which would consequently alter the orbital frequency of surrounding particles. In contrast, the lower spin value of $ a/M = 0.15 $ for XTE J1859+226 indicates a reduced dragging effect.

It is also shown in Table~\ref{Table3}, the best-fit values of orbital radii $r/M$ range from $5.68$ to $6.85$, suggesting that the QPOs phenomena occur near the horizon of the black hole. Specifically, for XTE J1859+226, we can see that the best-fit values of orbital radii $r/M = 6.85$, corresponding to a more distant orbital location where particles experience weaker gravitational influences, resulting in lower QPO frequencies. For GRO J1655-40 and H1743-322, we could see that the best-fit values of orbital radii $r/M \approx 5.68$, indicating that their particle orbits are closer to the horizon and are strongly affected by the gravitational field, leading to higher QPO frequencies.

In addition, Table~\ref{Table3} also shows that the lower limit of the parameter $h/M$ ranges from $-0.0509$ to $-0.2943$, reflecting the different effects of the Horndeski scalar field under different black holes. Particularly, $h/M > -0.0509$ of GRO J1655-40 indicates that the Horndeski correction effect in this system is minimal, similar to the classical Kerr solution in general relativity. While $h/M > -0.2943$ of H1743-322 indicates that the scalar field has a relatively significant effect on the motion of particles in its vicinity, which may lead to more obvious changes in the QPO frequency. Through the fitting values of the three signal sources in Table~\ref{Table3}, we can find that the best value of $h/M$ comes from GRO J1655-40. At the $95 \%$ confidence level, the lower limit of $h/M$ is $-0.0509$.

It is noticed that \cite{Donmez:2024lfi} and \cite{Donmez:2024gqb} investigate the shock cone formation and the excitation of QPO frequencies resulting from the interaction between accretion flow and Horndeski black holes, utilizing full numerical simulations of Bondi-Hoyle-Lyttleton accretion. 
According to the numerical results of \cite{Donmez:2024gqb}, for a non-rotating Horndeski black hole, maintaining stable QPOs generally requires the parameter $h/M $to be greater than approximately  $-0.5$.  \cite{Donmez:2024lfi} indicates that, in rotating cases, the allowable range for $h/M$ depends on the spin parameter, and overly negative values of $h/M$ destabilize the shock cone structure, leading to the disappearance of QPO modes. Compared to our work, it is clear that we impose stronger constraints on $h/M$. Additionally, our study illustrates the differences in Horndeski corrections under various physical environments, providing robust evidence for verifying the applicability of Horndeski theory under various conditions of strong gravitational fields.

\section{Conclusion}\label{sec:conclusions}
In this paper, we have presented the constraints placed by QPOs on the parameters of Horndeski rotating black holes observed in three X-ray binaries, namely GRO J1655-40, XTE J1859+226, and H1743-322. We constrained the parameters for the mass, spin, orbital radius of the black hole, and parameters of the Horndeski scalar field using MCMC simulations.

Our results show that with an increase in the spin parameter $a/M$, both  $\nu_\phi$  and $\nu_\theta$  decrease, with a more pronounced decrease in $\nu_\theta$, while $\nu_r$ shifts to smaller $r/M$. On the other hand, increasing  $h/M $reduces $\nu_\phi$ and $\nu_\theta$ in smaller $r/M$, but increases them in larger $r/M$. The $ \nu_\phi$ curves at different $h/M$ intersect at a specific radius, highlighting a transition in scalar field effects.

Moreover, the MCMC analysis revealed measurable deviations from the  Kerr black hole due to the presence of the Horndeski scalar hair. The lower bounds of the Horndeski hair parameter vary across different systems, with the weakest constraint found in GRO J1655-40 ($h/M > -0.0509$), indicating that this black hole closely resembles a classical Kerr black hole. In contrast, the strongest deviation was observed in H1743-322 ($h/M > -0.2943$), implying a more significant influence of the scalar field in this system. Nevertheless, our constraint on the Horndeski hair is much more stricter than QPOs simulation from other accretion models \cite{Donmez:2024gqb}. 

Our work demonstrates the power of QPOs as a   test of modified gravity theories. The ensuing parameter constraints not only increase the astrophysical understanding of spinning black holes but also provide valuable insights into the phenomenology of scalar-tensor gravity and the deviations from general relativity. Future research can extend this work by analyzing a larger data set of QPO data for stellar-mass and supermassive black holes, testing other modified gravity theories, and other signatures such as black hole shadows, gravitational waveforms, and continuum fits. Multi-messenger approaches will be vital for testing the robustness of the Horndeski and other alternatives of gravity around the horizon of compact objects, especially testifying the no hair theorem of black hole.

\section*{Acknowledgments}
This work is partly supported by Natural Science Foundation of China under Grants No.12375054 and 12405067.  
Meng-He Wu is also sponsored by Natural Science Foundation of Sichuan (No. 2025ZNSFSC0876).
H.G. is supported by the Institute for Basic Science (Grant No. IBS-R018-Y1).

\bibliographystyle{JHEP}
\bibliography{refers}

\end{document}